\documentclass[showpacs,preprintnumbers,amsmath,amssymb,twocolumn,superscriptaddress,pra]{revtex4}
\usepackage{times}
\usepackage{graphicx}
\usepackage{amsfonts}
\usepackage{amsmath, amsthm, amssymb}
\usepackage{dsfont}

\newcommand{\tom}{\tilde{\omega}_{x,\alpha}}

\newcommand{\vk}{{\boldsymbol{k}}} 
\newcommand{\vecr}{\boldsymbol{r}} 
\newcommand{\vR}{\boldsymbol{R}} 
\newcommand{\vp}{\boldsymbol{p}}

\newcommand{\e}[1]{\mathrm{e}^{#1}}

\newcommand{\vd}{\boldsymbol{\delta}}
 
\newcommand{\eg}{\textit{e.g. }}

\def\i{\mathrm{i}}

\begin{document}
\title[Calculation of Drag and Superfluid Velocity from the Microscopic Parameters and Excitation Energies of a  Two-Component Bose-Einstein Condensate on an Optical Lattice]
{Calculation of Drag and Superfluid Velocity from the Microscopic Parameters and Excitation Energies of a  Two-Component Bose-Einstein Condensate on an Optical Lattice}
\author{Jacob Linder}
\affiliation{Department of Physics, Norwegian University of
Science and Technology, N-7491 Trondheim, Norway}
\author{Asle Sudb{\o}}
\affiliation{Department of Physics, Norwegian University of
Science and Technology, N-7491 Trondheim, Norway}

\date{Received \today}
\begin{abstract}
We investigate a model of a two-component Bose-Einstein condensate residing on an optical lattice. Within a Bogolioubov-approach at the mean-field level, we derive exact analytical expressions for the 
excitation spectrum of the two-component condensate when taking into account hopping and interactions between arbitrary 
sites. Our results thus constitute a basis for works that seek to clarify the effects of higher-order interactions in 
the system. We investigate the excitation spectrum and the two branches of superfluid velocity in more detail for two 
limiting cases of particular relevance. Moreover, we relate the hopping and interaction parameters in the effective Bose-Hubbard model to microscopic parameters in the system, such as the laserlight wavelength and atomic masses of the components in the condensate. These results are then used to calculate analytically and numerically the drag coefficient between the components of the condensate. We find that the drag is most effective close to the symmetric case of equal masses between the components, regardless of the strength of the intercomponent interaction and the lattice well depth.
\end{abstract}

\maketitle

\section{Introduction}

The emergence of laser cooling techniques and their applications to realizing the phenomenon of Bose-Einstein condensation 
(BEC) in the laboratory, has paved the way for a study of the rich physics present when atoms condense at ultralow 
temperatures on an optical lattice \cite{dalfovo_rmp_99, leggett_rmp_01,morsch_rmp_06}. The BEC itself 
is a coherent matter wave, and has attracted much attention both theoretically and experimentally over the past decade. One 
of the remarkable features of a BEC residing on an optical lattice is the extent to which physical quantities such 
as tunnel coupling and on-site interaction may be controlled experimentally simply by adjusting the lattice parameters. This 
is done by controlling the interference pattern of the lasers setting up the optical lattice. For instance, by causing the 
depth of the lattice potential to increase, when would expect a resultant decrease of the hopping amplitudes and an increase 
of the on-site interaction. The possibility to alter the lattice parameters directly during the experiment, and thus 
influencing the physics, is clearly intriguing. Moreover, experiments carried out on such systems
are extremely well controlled since there is no disorder present. Since the atoms reside on an enginereed 
lattice, it is possible to investigate the physics by means of standard theories in condensed matter physics, such as 
the Bose-Hubbard model \cite{jaksch_prl_98}. As pointed out in Ref. \cite{morsch_rmp_06}, BECs residing on optical 
lattices have several advantages compared to ultracold atoms in a non-condensed phase. The main point is that the 
temperatures and densities for ultracold atoms and BECs both differ by three to four orders of magnitude. One 
consequence of the much higher particle densities for BECs is that atomic interactions become crucial with regard 
to the physics.  
\par
By allowing for more than one component of bosonic atoms on an optical lattice, one opens up an exciting avenue of physics to 
explore \cite{mazzarella_pra_06, chen_pra_03,fil_pra_05,kaurov_prl_05, dahl_prb_08,dahl_prl_09}. The physical realization of 
such a multicomponent BEC includes condensates with spin degrees of freedom (spinor condensates), two or more hyperfine states 
of the same atomic species that condense simultaneously, or simply two distinct atomic species. The two-component condensate 
has been shown to be a more rich environment to explore than a single-component BEC due to the possibility of an ``entrainment" 
coupling between the condensate components, see Ref. \onlinecite{dahl_prb_08,dahl_prl_09} and references therein. Such a system 
may be studied at a mean-field level by employing a Bogolioubov-approach, which may provide information about both the transition 
from a superfluid to Mott insulating state and also the quasiparticle excitation energies which arise from the condensate. By 
means of the Landau criterion, it is also possible to obtain information about the superfluid velocity of the condensate from 
the excitation spectrum. 

\begin{figure}[b!]
\centering
\resizebox{0.3\textwidth}{!}{
\includegraphics{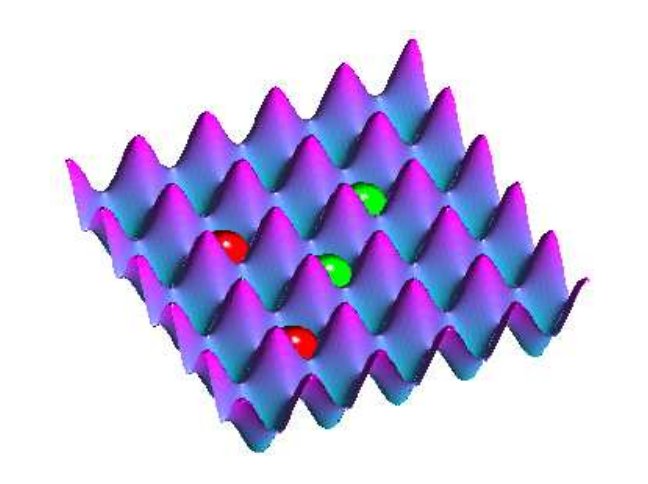}}
\caption{(Color online) An optical lattice setup by counter-propagating lasers serves as a potential landscape for two atomic species, 
denoted by the red (dark) and green (light) spheres. Each species of atoms may hop from site to site and also interact with both inter- and intraspecies 
atoms.}
\label{fig:lattice}
\end{figure}

Very recently, the excitation spectrum for a two-component Bose-Einstein condensate was obtained for a limiting case in 
Ref. \cite{liu_pra_07}. In that work, the author presented a correction to erroneous results previously reported in the 
literature \cite{gu_pla_05}. The calculations were performed under the standard assumptions of nearest-neighbor hopping 
and on-site interactions only. It would clearly be of interest extend calculations beyond these approximations, in order to 
investigate how the excitation spectrum is affected by taking into account longer-range hopping and longer-range
interactions. One of the purposes of the present paper is to extend the calculations of Ref. \onlinecite{liu_pra_07}
in this direction.
\par
Another goal in this paper is to address the effect of drag between the atomic components in a two-component Bose-Einstein condensate residing on an optical lattice. Such a drag effect points to a mutual transfer of motion between the components, and was first investigated in $^3$He-$^4$He superfluid mixtures by Andreev and Bashkin \cite{andreev_bashkin}. In Ref. \cite{fil_pra_05}, the drag effect for a two-component Bose gas was explored in the continuum limit. We will here derive an analytical expression for the intercomponent drag $\rho_d$ in a two-component Bose-Einstein condensate residing on an optical lattice, and relating it directly to the microscopic parameters in the system which are possible to tune experimentally.
\par
We organize this paper as follows. In Sec. \ref{sec:theory}, we establish the theoretical framework to be used in deriving our main 
results. In Sec. \ref{sec:results}, we provide an analytical solution for the excitation spectrum of a two-component Bose-Einstein 
condensate for arbitrary hopping and interaction between sites (Sec. \ref{sec:general}) and investigate the superfluid velocity 
and phase-separation condition in more detail for two limiting cases in Sec. \ref{sec:caseI} and \ref{sec:caseII}. Also, we present a correction to the condition for phase-stability of the two 
components, which determines whether the species are spatially miscible or not. In Sec. \ref{sec:drag}, we first relate analytically the 
parameters in the two-component Bose-Hubbard model directly to the fundamental physical quantities such as mass and trapping potential. Then, we combine these results with the expressions for the excitation energies in Sec. \ref{sec:caseI} and obtain an analytical equation for the drag coefficient in the system. The drag coefficient is then studied as a function of the microscopic parameters. Finally, we give concluding 
remarks in Sec. \ref{sec:summary}. The system under consideration is shown schematically 
in Fig. \ref{fig:lattice}.

\section{Theory}\label{sec:theory}

The starting point for our calculations is a microscopic Hamiltonian $\hat{H}$ for an ensemble of bosonic atoms that are confined by 
a slowly varying external harmonic trapping potential $V_{\text{T},\alpha}(\vecr)$ and subject to an additional optical lattice 
potential $V_0(\vecr)$. In terms of boson field operators $\psi_\alpha(\vecr)$, where $\alpha$ denotes the boson-component, $\hat{H}$ 
may be written as $(\hbar=1)$
\begin{align}\label{eq:H1}
&\hat{H} = \sum_\alpha \int \text{d}\vecr \psi_\alpha^\dag(\vecr)\Big[-\frac{\nabla^2}{2m_\alpha} - \mu_\alpha + V_0(\vecr) + V_{\text{T},\alpha}(\vecr)\Big]\notag\\
&\times\psi_\alpha(\vecr) + \frac{1}{2} \sum_{\alpha\beta} \int \text{d}\vecr \psi^\dag_\alpha(\vecr)\psi^\dag_\beta(\vecr)\gamma_{\alpha\beta}\psi_\beta(\vecr)\psi_\alpha(\vecr),
\end{align}
where $\gamma_{\alpha\beta}$ denotes the onsite-interaction for both boson species, $m_\alpha$ is the mass
of boson species $\alpha$, and $\mu_\alpha$ is its chemical potential. Specifically, we have \cite{fil_pra_05}
\begin{displaymath}
\gamma_{\alpha\beta} = \left\{ \begin{array}{ll}
4\pi a_\alpha/m_\alpha & \textrm{if $\alpha=\beta$}\\
2\pi(m_A+m_B)a_{AB}/m_Am_B & \textrm{if $\alpha\neq\beta$}
\end{array} \right.
\end{displaymath}
Here, $a_\alpha,a_{AB}$ are intraspecies and interspecies $s$-wave scattering lengths. 
The interaction strength is assumed to be repulsive and, in general, different for each of the boson components: $\{a_\alpha,a_{AB}\}>0$. To 
obtain a second-quantized Hamiltonian in a lattice-formulation, we assume that the field operators $\psi_\alpha(\vecr)$ may be expanded in 
a Wannier function basis set. The physical motivation for this is that the bosons are assumed to spend most of their time in the minima of 
the optical lattice potential, with occasional tunneling from one site to another. In this case, a set of localized Wannier functions 
where only the lowest lying excitation level is taken into account is expected to be a reasonable choice of basis. We consider here a 
two-dimensional model, such that $\psi_\alpha(\vecr) = \sum_i b_{i\alpha} w_\alpha(x-x_i)w_\alpha(y-y_i)$. Here, $b_{i\alpha}$ are boson 
annihilation operators for species $\alpha$ on the lattice point $i$, while $W_\alpha(\vecr-\vecr_i)=w_\alpha(x-x_i)w_\alpha(y-y_i)$ are
single-particle Wannier states for boson species $\alpha$ centred around lattice point $i$ at $(x_i,y_i)$. Inserting this expansion 
into Eq. (\ref{eq:H1}) yields an effective Bose-Hubbard like model, defined  by the Hamiltonian
\begin{align}\label{eq:H2} 
\hat{H} &= -\sum_\alpha \sum_{i\neq j} t_{ij\alpha} b_{i\alpha}^\dag b_{j\alpha} + \sum_{i\alpha} \varepsilon_{i\alpha} b_{i\alpha}^\dag b_{i\alpha} \notag\\
&+ \frac{1}{2} \sum_{ijkl} \sum_{\alpha\beta} U_{ijkl\alpha\beta} b_{i\alpha}^\dag b_{j\beta}^\dag b_{k\beta} b_{l\alpha}.
\end{align}
The parameters of this model are expressed as
\begin{align}\label{eq:parameters}
t_{ij\alpha} &= -\int \text{d}\vecr W_\alpha^*(\vecr-\vecr_i)\Big[ -\frac{\nabla^2}{2m_\alpha} + V_0(\vecr)\notag\\
&\hspace{1.2in}  + V_{\text{T},\alpha}(\vecr)\Big]W_\alpha(\vecr-\vecr_j),\notag\\
\varepsilon_{i\alpha} &= \int \text{d}\vecr W_\alpha^*(\vecr-\vecr_i)\Big[ -\frac{\nabla^2}{2m_\alpha} -\mu_\alpha+ V_0(\vecr) \notag\\
&\hspace{1.2in}  + V_{\text{T},\alpha}(\vecr) \Big]W_\alpha(\vecr-\vecr_i),\notag\\
U_{ijkl\alpha\beta} &= \gamma_{\alpha\beta}\int \text{d}\vecr W_\alpha^*(\vecr-\vecr_i)W_\beta^*(\vecr-\vecr_j)\notag\\
&\hspace{0.55in} \times W_\beta(\vecr-\vecr_k)W_\alpha(\vecr-\vecr_l).
\end{align}
So far, we have made no approximations apart from the assumed field expansion. The integrals given above may be evaluated 
analytically by specifying the explicit form of $W_\alpha(\vecr)$. Let us consider the following generic form for the 
trap and laser potential:
\begin{align}
V_{\text{T},\alpha}(\vecr) &= \frac{m_\alpha}{2} (\omega_x^2x^2+\omega_y^2y^2+\omega_zz^2),\notag\\
V_0(\vecr) &= V_x\sin^2(k_xx) + V_y\sin^2(k_yy)+V_z\sin^2(k_zz).
\end{align}
Here, $\omega_j$ is the frequency of harmonic trapping potential associated with the $j$-direction while the wave vector $k_j$ for the optical lattice is related to the wavelength $\lambda$ of the laser light as $k_j = 2\pi/\lambda_j$, such that the lattice period becomes $a_j=\lambda_j/2$, $j\in\{x,y,z\}$.
In the harmonic approximation \cite{jaksch_prl_98, giampaolo_pra_04}, where the bosons have a small probability of being located far from each lattice site and higher energy states in each lattice potential may be neglected, the exact Wannier functions 
can be replaced with their harmonic-oscillator approximation to a satisfactory degree. Then, one may write
\begin{align}\label{eq:frequency}
w_\alpha(x-x_i) &= \Big(\frac{m_\alpha \tom }{\pi}\Big)^{1/4}\e{-m_\alpha (x-x_i)^2/2},\notag\\
\tom &= \sqrt{\omega_x^2 + 2V_xk_x^2/m_\alpha}.
\end{align}
and similarly for $w_\alpha(y-y_i)$ and $w_\alpha(z-z_i)$. Since the Wannier functions are known, one may derive analytical expressions that relate 
the parameters in Eq. (\ref{eq:H2}) to the microscopic parameters in the system. For the hopping term $t_{ij,\alpha}$, previous works have neglected the influence of the trapping potential on this parameter by demanding that $V_{\text{T},\alpha}(\vecr)$ varies much more slowly than $V_0(\vecr)$. In this work, we derive a more general expression for both the hopping parameter and the interaction term by generalizing previous results to the two-component case and also by including the effect of the trapping potential. This is done towards the end of 
Sec. \ref{sec:results} 

\section{Results}\label{sec:results}

We now proceed to derive an analytical expression for the excitation energies of the elementary quasiparticles of the condensate. 
The standard approximation consists of only considering nearest-neighbor hopping and on-site interactions. To begin with, we 
include all orders of hopping and interactions without any site-limitation. We then explicitly consider two cases of 
particular relevance. Finally, we relate the microscopic parameters of the system to the hopping and interaction term in the effective Bose-Hubbard Hamiltonian.

\subsection{General solution}\label{sec:general}
By introducing a mean-field decomposition of the interaction terms allows us to consider the case where a macroscopic 
number of particles have condensed into the zero-momentum state. Let us define the Fourier-transformed boson operators
\begin{equation}
b_{i\alpha} = \frac{1}{\sqrt{N_s}} \sum_\vk b_{\vk\alpha}\e{-i\vk\mathbf{r}_i},
\end{equation}
which inserted into Eq. (\ref{eq:H2}) may be written as
\begin{align}\label{eq:hg1}
H &= \sum_{\vk\alpha} (\varepsilon_{\vk,\alpha} + T_\alpha)b_{\vk\alpha}^\dag b_{\vk\alpha} + \frac{1}{N_s}\sum_{\{\vk_i\}}\Big[ \sum_\alpha \frac{1}{2}\tilde{U}_\alpha(\vk_2,\vk_3,\vk_4) \notag\\
&\times b_{\vk_1\alpha}^\dag b_{\vk_2\alpha}^\dag b_{\vk_3\alpha} b_{\vk_4\alpha} \delta_{\vk_1+\vk_2,\vk_3+\vk_4} + \tilde{U}_{AB}(\vk_1,\vk_2,\vk_3,\vk_4) \notag\\
&\times b_{\vk_1 A}^\dag b_{\vk_2 A} b_{\vk_3 B}^\dag b_{\vk_4 B} \delta_{\vk_1+\vk_3,\vk_2+\vk_4}\Big],
\end{align}
where we have defined the generalized intraspecies potential 
\begin{align}\label{eq:ualpha}
\tilde{U}_\alpha(\vk_2,\vk_3,\vk_4) &= U_\alpha(0,0,0) + \sum_{\{\boldsymbol{\delta}_i\}} U_\alpha(\vd_1,\vd_2,\vd_3)\notag\\
&\times \e{\i(\vk_2\cdot\vd_1 - \vk_3\cdot\vd_2 - \vk_4\cdot\vd_3)},  
\end{align}
and the interspecies potential
\begin{align}\label{eq:uab}
\tilde{U}_{AB}&(\vk_1, \vk_2,\vk_3,\vk_4) = U_{AB}(0,0,0) + \sum_{\{\boldsymbol{\delta}_i\}} U_{AB}(\vd_1,\vd_2,\vd_3)\notag\\
&\times(\e{\i(\vk_1\cdot\vd_1 - \vk_2\cdot\vd_2 - \vk_4\cdot\vd_3)} + \e{\i(\vk_3\cdot\vd_1 - \vk_4\cdot\vd_2 - \vk_2\cdot\vd_3)}).
\end{align}
Above, the quantities $U_\alpha(\vd_1,\vd_2,\vd_3)$ and $U_{AB}(\vd_1,\vd_2,\vd_3)$ denote the interaction strengths and their dependence on the site distance between the particles involved in the scattering process, while $N_s$ denotes the number of lattice sites. Also, we have assumed that the energy off-set at each lattice site is simply a constant $\varepsilon_{i\alpha} = T_\alpha$. The interactions are related to the scattering potential $U_{ijkl\alpha\beta}$ as follows
\begin{align}
U_\alpha(\vd_1,\vd_2,\vd_3) = U_{i,i+\vd_1,i+\vd_2,i+\vd_3,\alpha\alpha},\notag\\
U_{AB}(\vd_1,\vd_2,\vd_3) = U_{i,i+\vd_1,i+\vd_2,i+\vd_3,AB},\notag\\
\end{align}
and are thus assumed to be independent on at which particular lattice site $i$ the scattering takes place, as is reasonable. The kinetic energy term is given by
\begin{align}
\varepsilon_{\vk,\alpha} = -\sum_{\vd} t_\alpha(\vd)\e{-\i\vk\cdot\vd},
\end{align}
where the summation over $\vd$ is to be taken over all neighbor sites. 
In Eqs. (\ref{eq:ualpha}) and (\ref{eq:uab}), the summation over $\{\vd_i\} = (\vd_1,\vd_2,\vd_3)$ is to be taken over all possible combinations of on-site and off-site lattice points except for pure on-site scattering $\{\vd_i\}=\{\boldsymbol{0}\}$. In this way, the first term in Eqs. (\ref{eq:ualpha}) and (\ref{eq:uab}) represents the on-site interaction while the second term incorporates scattering involving multiple sites.
\par
Since we are considering the condensed phase, we may write
\begin{align}\label{eq:nzero}
b_{0\alpha}b_{0\alpha}^\dag = b_{0\alpha}^\dag b_{0\alpha} + 1 \simeq N_{0\alpha} \gg 1,\notag\\
N_{0\alpha} = N_\alpha - \sum_{\vk}' b_{\vk\alpha}^\dag b_{\vk\alpha},
\end{align}
where the $'$ superscript over the sum denotes summation over all modes except $\vk=0$. 
Physically, we are stating that the number of atoms in the zero-mode state $\vk=0$ dominates the contribution to the total number of atoms for all $\vk$-modes. The biquadratic terms may be reduced to bilinear form by retaining only the interaction between the $\vk=0$ modes and other $\vk\neq0$ modes. Since the number of atoms $N_{0\alpha}$ in the $\vk=0$ mode for atom species $\alpha$ is assumed to satisfy Eq. (\ref{eq:nzero}), we may replace $b_{0\alpha}=b_{0\alpha}^\dag = \sqrt{N_{0\alpha}}$. 
\par
\begin{figure}[b!]
\centering
\resizebox{0.45\textwidth}{!}{
\includegraphics{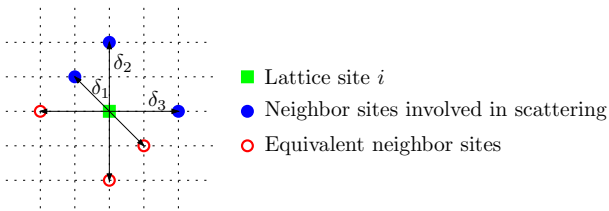}}
\caption{(Color online) The physical scenario expressed by Eq. (\ref{eq:criterion}).}
\label{fig:model}
\end{figure}

Next, we explicitly take into account the $\delta$-function constraints on the particle momenta in Eq. (\ref{eq:hg1}), which allows us to reduce the Hamiltonian to a sum over the atom species $\alpha$ and a single sum over momentum $\vk$.  In this way, one obtains
\begin{align}\label{eq:hel}
H = H_0 + \sum_\vk' \Big[H_{AB} + \sum_\alpha H_\alpha\Big],
\end{align}
where we have defined
\begin{align}
H_0 &= \sum_\alpha \Big[N_\alpha(T_\alpha+\varepsilon_{0,\alpha}) + \frac{N_\alpha^2}{2N_s} \tilde{U}_\alpha(0,0,0)\Big] \notag\\
&+ \frac{N_AN_B}{N_s} \tilde{U}_{AB}(0,0,0,0)
\end{align}
and the interaction terms
\begin{widetext}
\begin{align}
H_\alpha &= \epsilon_\vk^\alpha + (n_\alpha/2)[ \tilde{U}_\alpha(\vk,0,\vk) +  \tilde{U}_\alpha(0,\vk,0) +  \tilde{U}_\alpha(0,0,\vk) + \tilde{U}_\alpha(\vk,\vk,0) - 2\tilde{U}_\alpha(0,0,0) ]   b_{\vk\alpha}^\dag b_{\vk\alpha} \notag\\
&+ (n_\alpha/2)[\tilde{U}_\alpha(-\vk,0,0)b_{\vk\alpha}^\dag b_{-\vk\alpha}^\dag + \tilde{U}_\alpha(0,\vk,-\vk)b_{\vk\alpha} b_{-\vk\alpha}] \\
\text{ }\notag\\
H_{AB} &=  [ \tilde{U}_{AB}(\vk,0,0,\vk) b_{\vk A}^\dag b_{\vk B} + \tilde{U}_{AB}(0,\vk,\vk,0) b_{\vk A} b_{\vk B}^\dag + \tilde{U}_{AB}(\vk,0,-\vk,0) b_{\vk A}^\dag b_{-\vk B}^\dag + \tilde{U}_{AB}(0,\vk,-\vk,0) b_{\vk A} b_{-\vk B}] \notag\\
&\times\sqrt{n_An_B} + n_A[\tilde{U}_{AB}(0,0,\vk,\vk) - \tilde{U}_{AB}(0,0,0,0)] b_{\vk B}^\dag b_{\vk B} + n_B [\tilde{U}_{AB}(\vk,\vk,0,0) - \tilde{U}_{AB}(0,0,0,0)]b_{\vk A}^\dag b_{\vk A},
\end{align}
where $\epsilon_\vk^\alpha = \varepsilon_{\vk\alpha} + \sum_{\boldsymbol{\delta}} t_\alpha(\boldsymbol{\delta})$.
The above equation describes the Hamiltonian of a two-component Bose-Einstein condensate residing on an optical lattice with a drag between the atomic species. By diagonalizing Eq. (\ref{eq:hel}), we obtain the quasiparticle spectrum which allows for a further study of the different phases that may be expected for the condensate and also how the superfluid velocity depends on the interaction parameters.
Using the basis 
\begin{align}
\phi_\vk = [b_{\vk A}, b_{-\vk A}, b_{\vk B}, b_{-\vk B}, b^\dag_{\vk A}, b^\dag_{-\vk A}, b^\dag_{\vk B}, b^\dag_{-\vk B}]^\text{T},
\end{align}
the Hamiltonian can now be written in compact matrix form: 
\begin{align}\label{eq:hageneral}
H = H_0 + \frac{1}{4}\sum_\vk' \phi_\vk^\dag \check{\mathcal{M}}_\vk \phi_\vk,
\end{align}
where the matrix $\check{\mathcal{M}}_\vk$ reads
\begin{align}\label{eq:matgeneral1}
\check{\mathcal{M}}_\vk &= \begin{pmatrix}
\hat{M}_1(\vk) & \hat{M}_2(\vk) \\
\hat{M}_2(\vk)^* & \hat{M}_1(\vk)^*\\
\end{pmatrix},
\end{align}
upon defining the auxiliary matrices
\begin{align}\label{eq:matgeneral2}
\hat{M}_1(\vk) &= \begin{pmatrix}
E_A(\vk) & 0 & V_1(\vk) & 0\\
0 & E_A(-\vk) & 0 & V_1(-\vk)\\
V_1^*(\vk) & 0 & E_B(\vk) & 0 \\
0 & V_1^*(-\vk) & 0 & E_B(-\vk)\\
\end{pmatrix},\notag\\
\hat{M}_2(\vk) &= \begin{pmatrix}
0 & U_A(\vk) & 0 & V_2^*(\vk) \\
U_A(\vk) & 0 & V_2^*(-\vk) & 0 \\
0 & V_2^*(-\vk) & 0 & U_B(\vk) \\
V_2^*(\vk) & 0 & U_B(\vk) & 0 \\
\end{pmatrix}.
\end{align}
We have introduced the following notation:
\begin{align}
E_A(\vk) = \epsilon_\vk^A &+ \frac{n_A}{2}[ \tilde{U}_A(\vk,0,\vk) +  \tilde{U}_A(0,\vk,0) +  \tilde{U}_A(0,0,\vk) + \tilde{U}_A(\vk,\vk,0) - 2\tilde{U}_A(0,0,0)] \notag\\
&+ n_B[\tilde{U}_{AB}(\vk,\vk,0,0) - \tilde{U}_{AB}(0,0,0,0)],\notag\\
E_B(\vk) = \epsilon_\vk^B &+ \frac{n_B}{2}[ \tilde{U}_B(\vk,0,\vk) +  \tilde{U}_B(0,\vk,0) +  \tilde{U}_B(0,0,\vk) + \tilde{U}_B(\vk,\vk,0) - 2\tilde{U}_B(0,0,0)] \notag\\
&+ n_A[\tilde{U}_{AB}(0,0,\vk,\vk) - \tilde{U}_{AB}(0,0,0,0)],\notag\\
U_j(\vk) &= n_j \tilde{U}_j(-\vk,0,0) \text{ with } j=A,B,\; V_1(\vk) = \sqrt{n_An_B}\tilde{U}_{AB}(\vk,0,0,\vk),\; V_2(\vk) = \sqrt{n_An_B}\tilde{U}_{AB}(0,\vk,-\vk,0).
\end{align}
In order to obtain Eqs. (\ref{eq:matgeneral1}) and (\ref{eq:matgeneral2}), we made use of the fact that the matrix $\check{\mathcal{M}}_\vk\check{\sigma}_3$ must be Hermitian, since the eigenvalues have to be real (see discussion below).
Our ultimate goal is to obtain a Hamiltonian that may be written as
\begin{align}
H = \tilde{H}_0 + \frac{1}{4}\sum_\vk' \Phi_\vk^\dag \check{\mathcal{D}}_\vk \Phi_\vk,
\end{align}
where the matrix $\check{\mathcal{D}}_\vk$ contains the excitation energies. Note that $\tilde{H}_0$ will in general be different from $H_0$. The new basis $\Phi_\vk$ is related to the old one $\phi_\vk$ through the diagonalization matrix $\check{T}_\vk$, and also satsifies the correct boson commutation relation: $\Phi_\vk = \check{T}_\vk^\dag \phi_\vk,\; \Phi_\vk\Phi_\vk^\dag - (\Phi_\vk^\dag \Phi_\vk)^\mathcal{T} = \check{\sigma}_3.$
From the requirement that the new basis also consists of boson operators, one finds that the relation $\check{T}_\vk^\dag \check{\sigma}_3 \check{T}_\vk = \check{\sigma}_3$ must be satisfied. From this, one may infer that
$(\check{\mathcal{M}}_\vk\check{\sigma}_3) = \check{T}_\vk(\check{D}_\vk \check{\sigma}_3) \check{T}_\vk^{-1},
$
which means that $\check{T}_\vk$ diagonalizes the matrix $(\check{\mathcal{M}}_\vk\check{\sigma}_3)$. The corresponding eigenvalues are contained in the matrix $\check{D}_\vk \check{\sigma}_3$, and may be determined by considering
$|\check{\mathcal{M}}_\vk \check{\sigma}_3 - \Lambda\check{1} | = 0$. Evaluating the above determinant yields four distinct eigenvalues $\Lambda_\vk = \pm \mathcal{E}_{\vk\sigma}$, $\sigma=\pm1$.
\par
Before carrying out the diagonalization procedure, it is advantadgeous to make a simplifying observation: if the interaction potential satisfies
\begin{align}\label{eq:criterion}
U_\alpha(\vd_1,\vd_2,\vd_3) = U_\alpha(-\vd_1,-\vd_2,-\vd_3),
\end{align}
and similarly for $\alpha\to AB$, one may verify directly that $\{U_A(\vk), U_B(\vk), V_1(\vk), V_2(\vk)\}$ in Eq. (\ref{eq:matgeneral2}) are all even under inversion of momentum, i.e. $\vk\to (-\vk)$. Physically, Eq. (\ref{eq:criterion}) expresses that the scattering potential for a set of lattice sites and the sites obtained upon a mirror transformation, as shown in Fig. \ref{fig:model}, which is the case \textit{e.g.} for a square lattice. In addition, one may verify that $\{U_A(\vk), U_B(\vk), V_1(\vk), V_2(\vk)\}$ must all be real quantities for the same reason.
\par
Thus, we are finally able to give an analytical expression for the excitation energies $\Lambda_\vk$ for a two-component Bose-Einstein condensate with drag when taking into account arbitrary hopping and interaction between arbitrary sites. We find that 
\begin{align}\label{eq:eigenmain1}
\mathcal{E}_{\vk\sigma} = \frac{1}{2}\Big[&2[E_B^2(\vk)+E_A^2(\vk)] + 4[V_1^2(\vk) - V_2^2(\vk)] - 2[U_A^2(\vk)+U_B^2(\vk)] + 2\sigma \sqrt{R(\vk)}\Big]^{1/2},
\end{align}
where we have introduced
\begin{align}\label{eq:eigenmain2}
R(\vk) &= 8[V_1^2(\vk)+V_2^2(\vk)][U_A(\vk)U_B(\vk)+E_A(\vk)E_B(\vk)] + 4[V_2^2(\vk)-V_1^2(\vk)][U_A^2(\vk)+U_B^2(\vk)-E_A^2(\vk)-E_B^2(\vk)]\notag\\
&- 16V_1(\vk)V_2(\vk)[E_A(\vk)U_B(\vk) + E_B(\vk)U_A(\vk)]+ [E_A^2(\vk) + U_B^2(\vk) - E_B^2(\vk) - U_A^2(\vk)]^2,
\end{align}
Eqs. (\ref{eq:eigenmain1}) and (\ref{eq:eigenmain2}) represent one of our key results in this paper. Since there is no restriction 
on the sites involved in the hopping and interaction, the $\vk$-dependence of the eigenvalues cannot be evaluated analytically in 
any straight-forward manner. However, the above closed analytical form for the excitation energies may serve as a basis for numerical 
investigations of the interaction between the two atomic species in the condensate. Below, we consider two limiting cases of 
particular relevance which allow  further instructive analytical insight.
\end{widetext}

\subsection{Limiting case I: Nearest-neighbor hopping + on-site interactions}\label{sec:caseI}
We find that the terms in the Hamiltonian Eq. (\ref{eq:hageneral}) may now be written as 
\begin{align}\label{eq:h0}
H_0 &= \sum_\alpha \Big[\frac{U_\alpha N_\alpha^2}{2N_s} + N_\alpha(\varepsilon_{0\alpha} + T_\alpha)\Big] + \frac{U_{AB}N_AN_B}{N_s},
\end{align}
and we have introduced the basis vector 
\begin{align}\label{eq:drag_eq1} 
\phi_\vk = [b_{\vk A}, b_{-\vk A}, b_{\vk B}, b_{-\vk B}, b^\dag_{\vk A}, b^\dag_{-\vk A}, b^\dag_{\vk B}, b^\dag_{-\vk B}]^\text{T},
\end{align}
where the 'T' superscript denotes the matrix transpose. The matrix $\check{\mathcal{M}}_\vk$ has an $8\times8$ structure, and reads
\begin{align}\label{eq:matrix}
\check{\mathcal{M}}_\vk &= \begin{pmatrix}
\hat{X}_\vk & \hat{Y}_\vk \\
\hat{Y}_\vk & \hat{X}_\vk \\
\end{pmatrix},
\end{align}
upon defining the auxiliary matrices:
\begin{align}\label{eq:drag_eq2}
\hat{X}_\vk &= \begin{pmatrix}
E_\vk^A & 0 & F_{AB} & 0  \\
0 & E_\vk^A & 0 & F_{AB}  \\
F_{AB} & 0 & E_\vk^B & 0 \\
0 & F_{AB} & 0 & E_\vk^B  \\
\end{pmatrix},\notag\\
\hat{Y}_\vk &= \begin{pmatrix}
0 & F_A & 0 & F_{AB}  \\
F_A & 0 & F_{AB} & 0  \\
0 & F_{AB} & 0 & F_B  \\
F_{AB} & 0 & F_B & 0  \\
\end{pmatrix}.
\end{align}
Upon introducing $N_\alpha/N_s = n_\alpha$, we may write $F_{AB} = U_{AB}\sqrt{n_An_B},$ $F_\alpha = U_\alpha n_\alpha$, and $\epsilon_\vk^\alpha = t_\alpha\sum_{\boldsymbol{\delta}} \Big(1-\e{-\i\vk\cdot\boldsymbol{\delta}}\Big),$ $ E_\vk^\alpha = \epsilon_\vk^\alpha +F_\alpha,\; \alpha=A,B.$
By undertaking a diagonalization procedure, one obtains the excitation spectrum for the condensed ground-state. Some care must be exercised in this procedure, as the new quasiparticle operators in the diagonalized basis must also satisfy the boson commutation relations. 
As discussed previously, it is the matrix $\check{\mathcal{M}}_\vk\check{\sigma}_3$ that must be diagonalized to obtain the quasiparticle excitation energies. Evaluating the above determinant yields four distinct eigenvalues $\Lambda_\vk = \pm \mathcal{E}_{\vk\sigma}$, $\sigma=\pm1$, where
\begin{align}\label{eq:eigenvalue}
&\mathcal{E}_{\vk\sigma} = \Big[  \frac{\epsilon_\vk^A(\epsilon_\vk^A+2F_A) + \epsilon_\vk^B(\epsilon_\vk^B+2F_B)}{2} + \frac{\sigma}{2}   \notag\\
&\times\sqrt{[\epsilon_\vk^A(\epsilon_\vk^A + 2F_A) - \epsilon_\vk^B(\epsilon_\vk^B+2F_B)]^2 + 16F_{AB}^2\epsilon_\vk^A\epsilon_\vk^B}\Big]^{1/2}.
\end{align}
Note that in the limit of two decoupled Bose-Einstein condensates ($F_{AB}=0$) which are identical ($F_A=F_B=F$, $t_A=t_B=t$), we regain the well-known single-component spectrum $\mathcal{E}_\vk = \sqrt{\epsilon_\vk(\epsilon_\vk + 2F)}.$
The matrix $\check{D}_\vk$ now contains the excitation spectrum and reads (the choice of the order of the eigenvalues is arbitrary) 
\begin{align}
\check{D}_\vk = \text{diag}(\hat{d}_\vk,\hat{d}_\vk),\; \hat{d}_\vk = \text{diag}(\mathcal{E}_{\vk+}, \mathcal{E}_{\vk-}, -\mathcal{E}_{\vk+}, -\mathcal{E}_{\vk-}).
\end{align}
\par
Some comments are in order at this point. First of all, a similar approach to the condensed phase of a two-component Bose-Einstein condensate 
has been undertaken in both Ref. \cite{gu_pla_05} and \cite{liu_pra_07}. However, the final answer for the diagonalized spectrum appears to 
be erroneous in Ref. \cite{gu_pla_05}, where the effect of the drag (interspecies coupling $U_{AB}$) was completely disregarded in the 
excitation spectrum. Our results agree with the ones obtained in Ref. \cite{liu_pra_07}. The zero-temperature phase diagram for a 
two-component Bose-Einstein condensate on an optical lattice was analytically constructed in Ref. \cite{chen_pra_03}. Moreover, it 
was pointed out in Ref. \cite{oosten_pra_01} that within the framework employed here (Bogolioubov approach) one is able to obtain 
the criteria that demarcates the transition from a superfluid to Mott-insulator state, but one is  \textit{not} able to find the 
manifestation of this phase transition in \eg a sharp drop of the condensate fraction. 
\par
We will now proceed to investigate the superfluid velocity in more detail. The hydrodynamic flow in a Bose-Einstein 
condensate, and thus the superfluid velocity, may be probed experimentally by stirring the condensate with, 
for instance, a blue-detuned laser beam as in Ref. \cite{onofrio_prl_00}. In the present case, we 
find two branches \footnote{It should be noted that since the bosons present reside on a lattice, the authors of cond-mat/0607098 suggested that the superfluid velocity should be multiplied by a factor $m/m^*$ where $m^*$ is the effective band mass. However, this merely corresponds to a constant prefactor which we do not consider in more detail here.}
\begin{align}
\mathbf{v}_\sigma &= \nabla_\vk \mathcal{E}_{\vk\sigma}|_{\vk\to0}.
\end{align}
Below, we consider the one-dimensional case to obtain analytically transparent results which should elucidate 
the basic physics. Straight-forward derivation leads to:
\begin{align}\label{eq:vc}
v_\sigma = \Big[&\sigma a^2\sqrt{(t_AF_A-t_BF_B)^2 + 4F_{AB}^2t_At_B}\notag\\
&+a^2(t_AF_A+t_BF_B)\Big]^{1/2}.
\end{align}
This is consistent with the sound-like spectrum of Eq. (\ref{eq:eigenvalue}) in the long-wavelength limit $k\to0$. 
Note how the superfluid velocity for each branch vanishes when the interaction parameters $U_\alpha,U_{AB}$ in the problem
are set to zero. Moreover, the superfluid velocity $\mathbf{v}_-$ vanishes  if one of the hopping matrix
elements $t_A$  or $t_B$ vanishes, in which case the interspecies interaction parameter $U_{AB}$ is not
relevant in the superfluid velocity, such that $\mathbf{v}_+$ reduces to the superfluid velocity of a one-component 
Bose-Hubbard model. It is also interesting to generalize Eq. (\ref{eq:eigenvalue}) to the case of particles moving 
in a continuum, by substituting
\begin{align}
\epsilon_\vk^\alpha \to \frac{k^2}{2m_\alpha},\; \alpha=A,B,
\end{align}
in which case the superfluid velocity takes the form
\begin{align}
v_\sigma = \Bigg[&\sigma \sqrt{\Big(\frac{F_A}{2m_A} - \frac{F_B}{2m_B}\Big)^2 + \frac{F_{AB}^2}{m_Am_B}}\notag\\
&+\frac{F_A}{2m_A} + \frac{F_B}{2m_B}\Bigg]^{1/2}.
\end{align}
Again, the result reduces to that of a one-component Bose-Hubbard model for the case where one of the species 
becomes immobile, i.e. either $m_A$ or $m_B$ becomes infinite, and the superfluid velocities vanish in the 
non-interacting case. In the continuum picture, we may also generalize Bogolioubovs argument for the behavior 
of the excitations in the short- and long-wavelength limit. The limit of  the long-wavelength linear sound-like 
spectrum  is roughly demarcated by a wavevector which gives equal magnitude for the kinetic and potential 
energy terms in the quasiparticle dispersion relation. For component $\alpha$, the crossover wavevector 
$k_{c,\alpha}$ to the linear regime is given by
\begin{align}
\frac{k_{c,\alpha}^2}{2m_\alpha} &= n_\alpha (U_\alpha + n_{\bar{\alpha}}U_{\alpha\bar{\alpha}})\notag\\
&\Rightarrow k_{c,\alpha} = \sqrt{2m_\alpha n_\alpha(U_\alpha + n_{\bar{\alpha}}U_{\alpha\bar{\alpha}})} \sim \frac{1}{\xi_\alpha}.
\end{align}
where $\bar{\alpha}$ denotes the other component in the condensate while $\xi_\alpha$ is the coherence length. The 
physical picture is then that the atoms of species $\alpha$ move as free particles on short length scales compared 
to $\xi_\alpha$, while they move collectively at large length scales compared to $\xi_\alpha$. Some other aspects 
of the superfluid velocity for a two-component condensate with an energy dispersion appropriate for the continuum 
were considered in Ref. \cite{kravchenko_jltp_08}.
\par
We now proceed to investigate in detail how the superfluid velocity Eq. (\ref{eq:vc}) depends on the kinetic and 
potential energy terms in the problem. As seen, $v_\sigma$ depends on the hopping parameters $t_\alpha$, the 
intraspecies interactions $F_\alpha$, and the interspecies interaction $F_{AB}$. Upon choosing the parameters, 
we must ensure that the excitation energies remain real, as required for a stable phase of two interacting 
atomic species. From Eq. (\ref{eq:eigenvalue}), one infers that the solution may become imaginary
if the interaction $\gamma_{AB}^2$ becomes sufficiently large. The criterion for a stable coexistent 
phase of the condensed phase for both atomic species reads \cite{chen_pra_03}
\begin{align}\label{eq:criterion2}
\gamma_A\gamma_B > \gamma_{AB}^2.
\end{align}
Let us first investigate how the two branches of the superfluid velocity depend on the interspecies coupling. It is 
convenient to rewrite Eq. (\ref{eq:vc}) in terms of dimensionless parameters as follows:
\begin{align}\label{eq:vccaseI}
v_\sigma &= \sqrt{\zeta_A+\zeta_B+\sigma\sqrt{(\zeta_A-\zeta_B)^2 + 4\rho \zeta_A\zeta_B}},\notag\\
\zeta_\alpha &= t_\alpha F_\alpha a^2,\; \rho = \frac{F_{AB}^2}{F_AF_B} = \frac{\gamma_{AB}^2}{\gamma_A \gamma_B}.
\end{align}
It is interesting to note that for a fixed value of $\rho$, the tunnel coupling amplitudes $t_\alpha$ and the interaction 
parameters $U_\alpha$ play the same role. The expression for the superfluid velocity remains the same under exchange of 
these two energy scales. The physical regime of the normalized interspecies coupling is now $\rho\in[0,1]$, as demanded 
by Eq. (\ref{eq:criterion2}). In Fig. \ref{fig:cv2}, we show how the superfluid velocities in the two branches $v_\pm$ 
depend on the parameters in the problem. We give results for values of $\rho$ ranging from a weak interatomic scattering 
strength $(\rho=0.1)$ to a strong interaction $(\rho=0.9)$. As seen, the individual branches are not very sensitive to 
the value of $\rho$, but the two branches themselves differ qualitatively in their dependence on the hopping amplitudes 
and the potential energy. In the case of two symmetric Bose-Einstein condensates $(\zeta_A=\zeta_B)=\zeta$, one obtains 
from Eq. (\ref{eq:vccaseI}) that
\begin{align}
v_\sigma = [2\zeta(1 + \sigma\sqrt{\rho})]^{1/2}.
\end{align} 
The most interesting aspect of Fig. \ref{fig:cv2} is that the $v_-$ branch vanishes as $\rho\to1$. This means that a 
very small rotation or stirring of the condensate will trigger the $\sigma=-$ branch to become a normal fluid when $\rho\to1$.
\begin{widetext}
\text{ }\\
\begin{figure}[h!]
\centering
\resizebox{0.95\textwidth}{!}{
\includegraphics{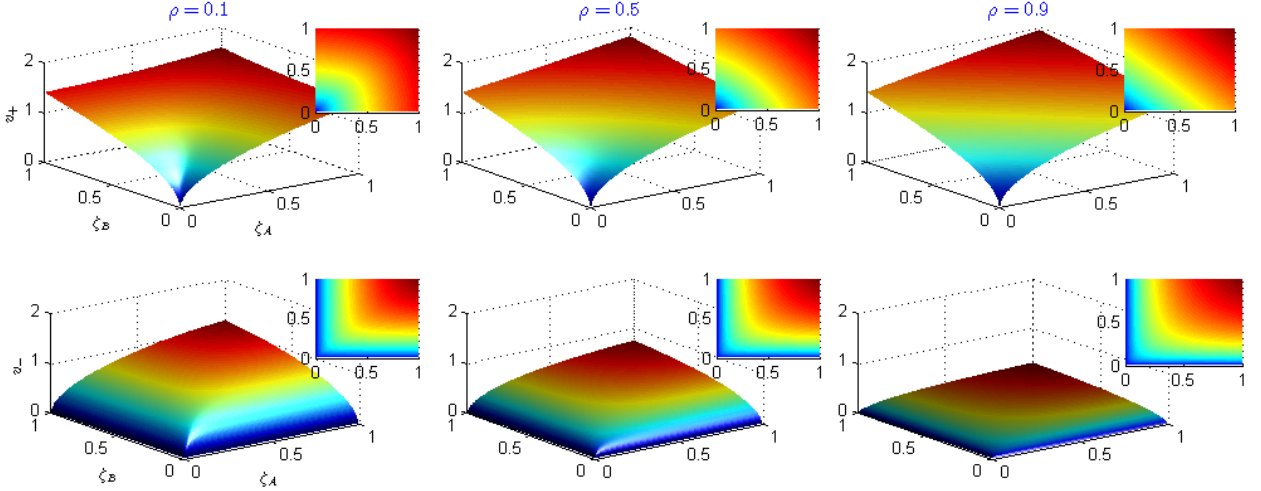}}
\caption{(Color online) Plot of the superfluid velocity for the two quasiparticle branches $v_\pm$. Above, we 
have used the dimensionless parameters $\zeta_\alpha = -t_\alpha F_\alpha a^2$ and 
$\rho = \gamma_{AB}^2/(\gamma_A \gamma_B)$. The latter is a measure for the interaction between 
the two atomic components on the condensate.}
\label{fig:cv2}
\end{figure}

\subsection{Limiting case II: Next nearest-neighbor hopping + off-site interactions}\label{sec:caseII}
We now go beyond the main approximation of Sec. \ref{sec:caseI} and allow additionally for both next 
nearest-neighbor hopping and nearest-neighbor interactions. In this way, the interaction term in Eq. 
(\ref{eq:H2}) becomes:
\begin{align}
\frac{1}{2} \sum_{ijkl} \sum_{\alpha\beta} U_{ijkl\alpha\beta} b_{i\alpha}^\dag b_{j\beta}^\dag b_{k\beta} b_{l\alpha} \to \frac{1}{2}\sum_{\alpha\beta} \Big[ \sum_i U_{\alpha\beta} b_{i\alpha}^\dag b_{i\beta}^\dag b_{i\beta} b_{i\alpha} + \sum_{i\neq j} U_{\alpha\beta}' n_{i\alpha}n_{j\beta}\Big],
\end{align}
\end{widetext}
where $U_{\alpha\beta}$ and $U_{\alpha\beta}'$ are the on-site and nearest-neighbor interactions, respectively. In order to obtain transparent analytical results, we consider the 1D case, corresponding to a trapping potential which is elongated in a "cigar"-like shape. Proceeding in an equivalent manner as in the previous sections, we finally obtain four distinct eigenvalues $\Lambda_\vk = \pm \mathcal{E}_{\vk\sigma}$, $\sigma=\pm1$, which are identical to Eq. (\ref{eq:eigenvalue}) except that $F_j\to F_\vk^j$, $j\in\{A,B,AB\}$, where we have defined the kinetic energy term
\begin{align}\label{eq:kin1}
\epsilon_\vk^\alpha = 2t_\alpha[1-\cos(ka)] + 2t_\alpha'[1-\cos(2ka)],\; \alpha=A,B
\end{align}
and the potential energy terms
\begin{align}
F_\vk^\alpha = n_\alpha[U_\alpha + 2U_\alpha'\cos(ka)],\; \alpha=A,B\notag\\
F_\vk^{AB} = \sqrt{n_An_B} [U_{AB} + 2U_{AB}'\cos(ka)].
\end{align}
In Eq. (\ref{eq:kin1}), $t_\alpha$ denotes the hopping parameter for nearest-neighbors while 
$t_\alpha'$ denotes the hopping parameter for next nearest-neighbors. One now obtains the two branches of superfluid velocities 
which may be expressed through dimensionless quantities as follows:
\begin{align}\label{eq:vccaseII}
v_\sigma &= \sqrt{\Psi_A+\Psi_B+\sigma\sqrt{(\Psi_A-\Psi_B)^2 + 4\nu \Psi_A\Psi_B}},
\end{align}
where we have defined
\begin{align}
\Psi_\alpha &= (t_\alpha + 4t_\alpha')(U_\alpha + 2U_\alpha')n_\alpha a^2,\notag\\
\nu &= \frac{(U_{AB}+2U_{AB}')^2}{(U_A + 2U_A')(U_B+2U_B')}.
\end{align}
Note that the above equations have exactly the same form as Eq. (\ref{eq:vccaseI}), and that one obtains 
$\nu\to\rho$, $\Psi_\alpha \to \zeta_\alpha$ in the limit $\{U_\alpha',U_{AB}\} \to 0$, as demanded by 
consistency. The stability condition for having a coexistent phase of the two superfluid branches is 
obtained by demanding that the eigenvalues are real, leading to the condition
\begin{align}
(\epsilon_\vk^A + 2F_\vk^A)(\epsilon_\vk^B + 2F_\vk^B) > 4(F_\vk^{AB})^2.
\end{align}
This is a generalization of the condition $U_AU_B > U_{AB}^2$ that arises from the standard assumption 
of only nearest-neighbor hopping and on-site interactions. Assuming $\epsilon_\vk^\alpha\geq0$, we may 
set $\epsilon_\vk^\alpha=0$ to find a more strict condition
\begin{align}
\frac{[U_{AB} + 2U_{AB}'\cos(ka)]^2}{[U_A+2U_A'\cos(ka)][U_B+2U_B'\cos(ka)]} > 1
\end{align}
for the phase-coexistence regime. Thus, for a strong repulsive interaction between the atomic species 
$A$ and $B$, one would expect that they do not coexist spatially but are instead separated into two 
distinct spatial regions. 

\subsection{Microscopic parameters and drag between superfluid components}\label{sec:drag}
We here derive explicit analytical expressions for the hopping and interaction parameters $t_\alpha$ and $U_{\alpha\beta}$ in our model. We will consider nearest-neighbor hopping and an optical lattice with intersite distance $a$. Our results are derived for the three-dimensional case, but are written down in a form which may be easily generalized to one or two dimensions.
\par
\begin{figure}[t!]
\centering
\resizebox{0.45\textwidth}{!}{
\includegraphics{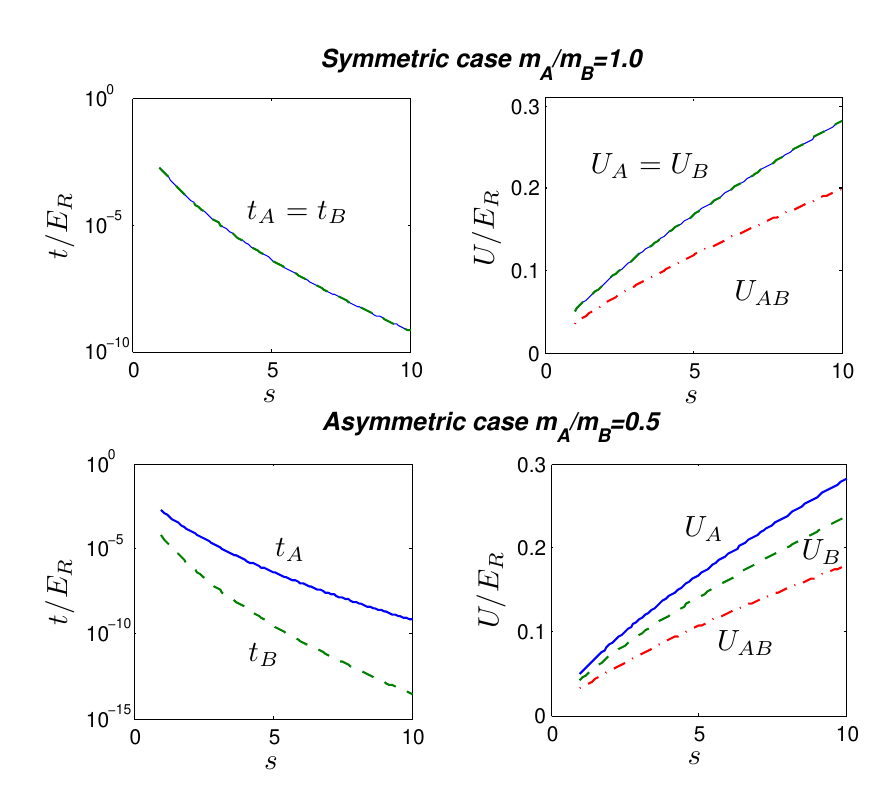}}
\caption{(Color online) Plot of the hopping and interaction parameters in the Bose-Hubbard model as a function of the trap depth $s=V_0/E_R$. }
\label{fig:parameters}
\end{figure}

Starting from the definitions in Eq. (\ref{eq:parameters}), we obtain
\begin{align}
U_{\alpha\beta} = \gamma_{\alpha\beta} \prod_j \sqrt{\frac{m_\alpha m_\beta \tilde{\omega}_{j\alpha} \tilde{\omega}_{j\beta}}{\pi(m_\alpha\tilde{\omega}_{j\alpha} + m_\beta\tilde{\omega}_{j\beta})}},
\end{align}
which is consistent with Eq. (9) in Ref. \cite{liu_pra_07}. The definition of $\tilde{\omega}_{j\alpha}$ was given in Eq. (\ref{eq:frequency}). Above, $j\in\{x,y,z\}$. Now, we present a derivation of the hopping term upon taking fully into account the trapping potential, which has been neglected in the literature so far. This is appropriate in a situation where the trapping potential has been turned off, allowing the condensate to expand very slowly. Inserting the potentials and the Wannier functions into Eq. (\ref{eq:parameters}), we obtain
\begin{align}
t_\alpha &= -\prod_j \sqrt{\frac{m_\alpha \tilde{\omega}_{j\alpha}}{\pi}} \int_{-\infty}^\infty d\vecr \e{-\vp_\alpha\cdot\tilde{\vR}(0)}\notag\\
&\times \Big[-\frac{\nabla^2}{2m_\alpha}+ \tilde{\boldsymbol{V}}_\alpha\cdot\tilde{\vR}(0)\Big] \e{-\vp_\alpha\cdot \tilde{\vR}(a)},
\end{align}
where we have defined
\begin{align}
\tilde{\vR}(a) &= \sum_j (j-a)^2 \boldsymbol{j},\; \vp_\alpha = (m_\alpha/2) \sum_j \tilde{\omega}_{j\alpha}\boldsymbol{j}\notag\\
\tilde{\boldsymbol{V}}_\alpha &= (m_\alpha/2) \sum_j (\omega_j^2 + 2V_jk_j^2/m_\alpha) \boldsymbol{j}.
\end{align}
After a shift of variables, we arrive at
\begin{align}
t_\alpha = -\prod_j \e{-(m_\alpha\tilde{\omega}_{j\alpha} a^2/4)}\sum_{j'} \tilde{\omega}_{j'\alpha}/2.
\end{align}
The ratio of the interaction term and the hopping term may now be evaluated straight-forwardly for any choice of microscopic parameters. Experimentally, it is possible to tune lattice parameters $V_0$ and $\lambda$ through the laserlight setting up the optical potential. Defining the atom recoil energy 
\begin{align}
E_{R,\alpha} = k^2/2m_\alpha = 2\pi^2/(m_\alpha\lambda^2),
\end{align}
one may then define the tunable parameter 
\begin{align}
s\equiv \frac{V_0}{E_{R,A}}
\end{align}
which captures the effect of both the lattice well depth $V_0$ and the lattice constant $a=\lambda/2$. We simply denote $E_{R,A} \equiv E_R$ from now on. For later use, we note that for a cubic lattice and in the absence of a trapping potential, we obtain the relations
\begin{align}\label{eq:bhparameters}
\frac{t_\alpha}{E_R} &= 3\e{-3\pi^2\sqrt{s m_\alpha/m_A}/4} \sqrt{s m_A/m_\alpha},\notag\\
\frac{U_\alpha}{E_R} &= \frac{a_\alpha}{\lambda} \frac{2}{\pi} \frac{m_A}{m_\alpha} ( 2\pi \sqrt{s m_\alpha/m_A})^{3/2},\notag\\
\frac{U_{AB}}{E_R} &= \frac{a_{AB}}{\lambda}\frac{1}{\pi} (1+m_A/m_B) \Big(\frac{4\pi\sqrt{s}}{1 + \sqrt{m_A/m_B}} \Big)^{3/2}.
\end{align}
upon choosing a positive sign for the hopping parameters. This fully determines the Bose-Hubbard parameters $t_\alpha$ and $U_{\alpha\beta}$ for a given set of microscopic parameters. To access the physically allowed regime of $a_{AB}$, we define
\begin{align}
\eta = \frac{\gamma^2_{AB}}{\gamma_A\gamma_B},\; \eta\in [0,1).
\end{align}
We plot in Fig. \ref{fig:parameters} the Bose-Hubbard parameters as a function of the lattice well depth $s$ for the case of symmetric $(m_A/m_B=1.0)$ and asymmetric $(m_A/m_B=0.5)$ two-component condensate. Moreover, $a_\alpha/\lambda = 10^{-3}$, corresponding to a scattering length of a few nm for a typical experiment. As seen, the hopping amplitude becomes comparable to the interaction term only for optical lattice potentials $V_0 \sim E_R$ yielding $s \sim 1$. We do not consider here very weak lattice potentials satisfying $s\ll1$, since the tight-binding model employed in the present paper no longer remains valid. 

As a practical application of our results for the excitation energies as well as the relation between the Bose-Hubbard parameters and microscopic parameters, we now study the magnitude of the intercomponent drag coefficient $\rho_d$ in a uniform two-component BEC. In particular, we investigate what values $\rho_d$ may realistically take for a relevant choice of microscopic parameters. The drag stems from a transfer of motion between the supercurrents for each component as a result of the interaction $\gamma_{AB}$, and vanishes in the case of two decoupled BECs. The free energy for a uniform two-component Bose-Einstein condensate may be written as \cite{andreev_bashkin}
\begin{align}
F = F_0 + V[\rho_A \boldsymbol{v}_A^2 + \rho_B\boldsymbol{v}_B^2 - \rho_d (\boldsymbol{v}_A - \boldsymbol{v}_B)^2]/2,
\end{align}
where $F_0$ contains terms independent of the superfluid velocities $\boldsymbol{v}_i$ for the two components and $V$ is the volume of the system. The terms $\rho_j$, $j\in A,B$ represent the mass densities of the two components. 
\par
In Ref. \cite{fil_pra_05}, an explicit expression was derived for the intercomponent drag $\rho_d$ for the case of small superfluid velocities (much smaller than the critical ones) in the continuum limit, i.e. with free-boson dispersion relations $\epsilon_{\vk}^\alpha = k^2/2m_\alpha$. However, the drag between components on an optical lattice remains to be investigated. In what follows, we shall calculate $\rho_d$ as a function of the microscopic parameters in the problem. This is accomplished by virtue of our analytical expressions for both the quasiparticle energies (Sec. \ref{sec:caseI}) and the parameters in the effective Bose-Hubbard Hamiltonian derived previously in this section. We here focus on the zero-temperature case, i.e. far away from the critical temperature, where our mean-field approach should be viable.
\par
We now derive an analytical expression for $\rho_d$ from the microscopic Hamiltonian determined by Eqs. (\ref{eq:hageneral}), (\ref{eq:h0}), (\ref{eq:drag_eq1}), (\ref{eq:matrix}), (\ref{eq:drag_eq2}). Our strategy is to let $\vk\to\vk - m_\alpha\boldsymbol{v}_\alpha$ in the Hamiltonian, leading to the Doppler-shifted energies
\begin{align}
\epsilon_\vk^\alpha \to \epsilon_\vk^\alpha - m_\alpha \boldsymbol{v}_\alpha \cdot \nabla_\vk \epsilon_\vk^\alpha.
\end{align}
The energy eigenvalues may then be solved by expanding the characteristic polynomial in orders of $\boldsymbol{v}_\alpha$, along the lines of \cite{herland_diplom}. At zero temperature, one obtains the following expression for the drag coefficient:
\begin{align}\label{eq:dragfinal}
\rho_d = \frac{4m_Am_Bt_At_B}{N_xN_yN_za} \sum_\vk' \frac{F_{AB}^2 \epsilon_\vk^A\epsilon_\vk^B \sin^2(k_xa)}{\mathcal{E}_{\vk,+}\mathcal{E}_{\vk,-}(\mathcal{E}_{\vk,+} + \mathcal{E}_{\vk,-})^3}.
\end{align}
Just like in the continuum limit treated in Ref. \cite{fil_pra_05}, we find that the drag coefficient is independent of the sign of the intercomponent scattering $F_{AB}$. It is also seen that Eq. (\ref{eq:dragfinal}) is always positive, $\rho_d>0$. Our results Eq. (\ref{eq:dragfinal}) may thus be considered as a generalization of the drag coefficient in Ref. \cite{fil_pra_05} to an optical lattice scenario. 
\par
\begin{figure}[h!]
\centering
\resizebox{0.48\textwidth}{!}{
\includegraphics{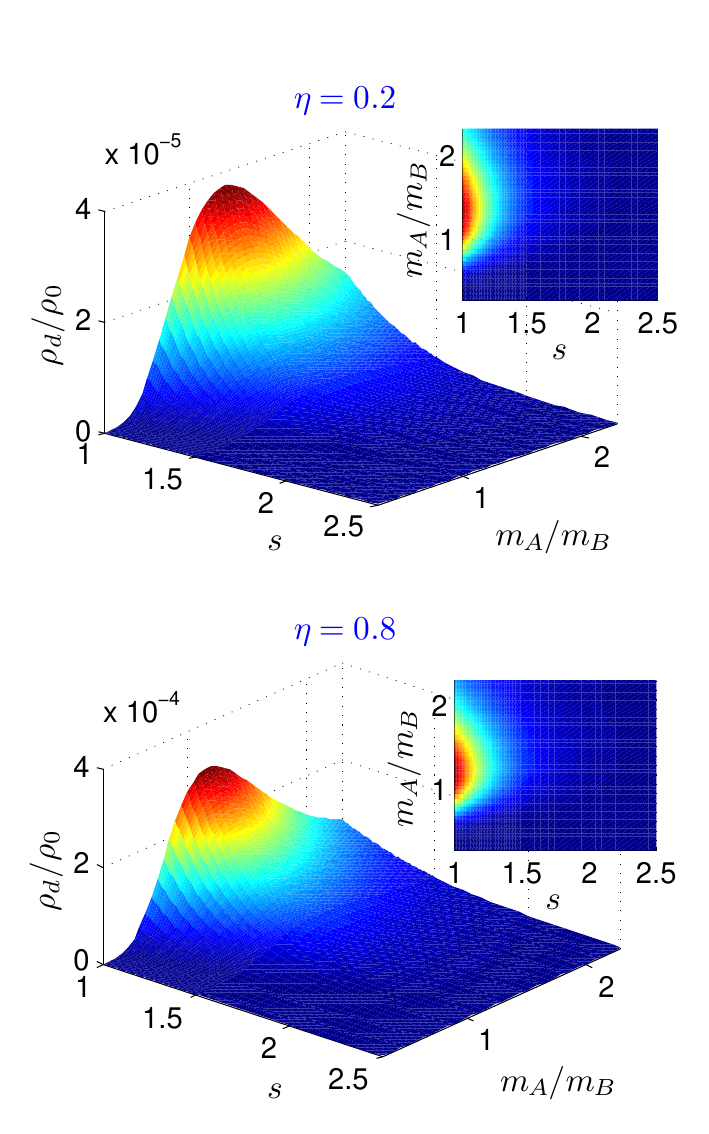}}
\caption{(Color online) Contour plot of the normalized drag coefficient $\rho_d/\rho_0$ as a function of the mass ratio $m_A/m_B$ and the lattice well depth $s$. Two different values of $\eta$ have been used, corresponding to weak ($\eta=0.2$) and strong ($\eta=0.8$) intercomponent scattering.}
\label{fig:contour_rho}
\end{figure}

We now proceed to investigate the behavior of the drag coefficient numerically on a $50\times50\times50$ cubic lattice $(N_j=50,\; V_j=V_0),\; j=\{x,y,z\}$, which corresponds to the experimental setup of Ref. \cite{greiner_nature_02}. Moreover, we fix $n_A=n_B=\sqrt{2}$, corresponding to an incommensurate filling as demanded for the superfluid phase. Let us define the normalized and dimensionless drag coefficient $\rho_d/\rho_0$, where $\rho_0 = m_A N_A/V$.
For a fixed intracomponent interaction strength $a_\alpha$, the drag coefficient will thus depend on the strength of the intercomponent scattering $\eta$, the mass ratio $m_A/m_B$, and the lattice well depth $s$. These microscopic parameters also determine the Bose-Hubbard parameters $t_\alpha$ and $U_{\alpha\beta}$ through Eq. (\ref{eq:bhparameters}). We present the dependence of $\rho_d/\rho_0$ on $m_A/m_B$ and $s$ in Fig. \ref{fig:contour_rho} for both a weak ($\eta=0.2$) and strong ($\eta=0.8$) intercomponent scattering.

Qualitatively, it is seen that the plots are similar. With increasing lattice well depth $s$, the drag coefficient quickly 
diminishes in size. It is interesting to note that the drag coefficient is at its largest for a mass ratio $m_A/m_B \sim 1$, 
regardless of the value of $s$. This suggests that the velocity-drag effect between the components becomes most efficient 
when they have similar masses, which is reasonable. If $\rho_d\neq0$, it is possible that the superfluid motion of one 
component induces a supercurrent in the other component purely by a drag effect, since the expressions for the 
supercurrents $\boldsymbol{j}_\alpha$ may be written as \cite{andreev_bashkin}:
\begin{align}
\label{currents}
\boldsymbol{j}_A = (\rho_A-\rho_d)\boldsymbol{v}_A + \rho_d\boldsymbol{v}_B,\; \boldsymbol{j}_B = (\rho_B-\rho_d)\boldsymbol{v}_B + \rho_d\boldsymbol{v}_A.
\end{align}
Thus, one may have $\boldsymbol{j}_\alpha\neq0$ even with $\boldsymbol{v}_\alpha=0$. 
\par
We end this section by briefly commenting on the positivity of $\rho_d$ that we find. In previous works on two-component 
Bose condensates, a negative drag has been found  numerically in Monte Carlo computations \cite{kaurov_prl_05}, and 
starting from such a negative value, highly unusual vortex states in a rotating multi-component Bose condensate, have 
been predicted which have no counterpart in the case $\rho_d > 0$ \cite{dahl_prl_09}. In particular, a superfluid 
bosonic density wave, corresponding to a vortex system which has the characteristics of a liquid and a solid at the same 
time, has been reported \cite{dahl_prl_09}. Therefore, a negative drag has extremly important ramifications for 
the physics of these systems. Physically, a positive drag coefficient means, by virtue of Eq. \ref{currents}, that 
a superfluid flow in one component of the condensate induces a co-directed, not counter-directed, flow in the
other component. It should be noted that the negative drag reported in Ref. \onlinecite{kaurov_prl_05} 
was obtained in a limit where the bosons on the optical lattice were very strongly interacting (essentially the hard-core 
boson limit) and the system was close to half-filling. Under such circumstances, one may expect a backflow of one species 
of bosons when a boson of one component hops from one lattice site to another site occupied by the other component. Our 
approach, on the other hand, is essentially a weak-coupling approach which cannot capture such physics, and this issue 
warrants further investigation using more suitable strong-coupling approaches.    

\section{Summary}\label{sec:summary}
In conclusion, we have investigated the excitation spectrum, superfluid velocity, and inter-component drag-coefficient 
for a two-component Bose-Einstein condensate on an optical lattice. We have derived analytical expressions for the 
excitation energies for arbitrary hopping and interaction between sites in Eqs. (\ref{eq:eigenmain1}) and 
(\ref{eq:eigenmain2}). We have investigated the excitation spectrum, superfluid velocity, and 
phase-separation condition in more detail for two important limiting cases of the general expressions. 
The critical superfluid velocity may be probed experimentally by using \eg a laser as a macroscopic 
object to stir the hydrodynamic flow in the two-component Bose-Einstein condensate. Moreover, we have 
derived an analytical expression for the drag coefficient between the components when the condensate 
resides on an optical lattice in a weak-coupling approximation and found that it is always positive. 
This means that drag induces co-directed flows of superfluid components, and not
counter-flows, in the weak-coupling limit. 
We find that the transfer of motion from one supercurrent to another 
becomes most efficient for a mass ratio close to one for the two components, regardless of the 
lattice well depth or the intercomponent scattering strength.

\section*{Acknowledgments}
\noindent E. Babaev, S. Kragset, and Z. Tesanovic are acknowledged for very useful discussions. 
E. K. Dahl, E. V. Herland, and M. Taillefumier are thanked for helpful input. This work was 
supported by the Research Council of Norway, Grants No. 158518/431 and No. 158547/431 (NANOMAT), 
and Grant No. 167498/V30 (STORFORSK).

\end{document}